\documentclass[a4paper,11pt]{article}
\pdfoutput=1 

\usepackage{jinstpub} 
\usepackage{siunitx}
\usepackage{lineno}

\title{The Radio Neutrino Observatory in Greenland (RNO-G) - System Design and Instrumentation}

\author[a,1]{N. Heyer\note{Corresponding author.}}

\affiliation[a]{Uppsala University, Dept. of Physics and Astronomy, Box 516, S-75120 Uppsala, Sweden}
\affiliation[b]{University of California, Department of Physics and Astronomy,
Irvine, CA,92617, USA}

\emailAdd{nils.heyer@physics.uu.se}

\abstract{The Radio Neutrino Observatory in Greenland (RNO-G) is designed to measure cosmic neutrinos in the EeV range through an array of radio antennas detecting nanosecond-long radio flashes. In this contribution, the RNO-G electronic system is outlined, illustrating how signals are measured from the antennas to the recorded pulses. We present new instrumentation developments, focusing on electric melters to install LPDA antennas into the Greenlandic ice sheet. The new system will simplify deployment and reduce positioning uncertainty to \SI{\pm 2}{\centi\meter} (x/y) and \SI{\pm 1}{\deg} (rotational). We report on field tests from the RNO-G site at Summit Station.}

\keywords{Neutrino detectors, Large detector systems for particle and astroparticle physics, Antennas, Detector design and construction technologies and materials}

\collaboration[c]{on behalf of the RNO-G collaboration\\ with S. Barwick$^{b}$ and A. Nielsen$^{b}$}

\proceeding{Topical Workshop on Electronics for Particle Physics - TWEPP2026\\
  28$^{\text{th}}$ of September - 2$^{\text{nd}}$ of October 2026\\
  Castelldefels, Barcelona, Spain}

\begin{document}
\maketitle
\flushbottom

\section{The Radio Neutrino Observatory in Greenland}
\label{sec:intro}

\begin{wrapfigure}{r}{0.55\textwidth}
         \centering
         \includegraphics[height=4.2in]{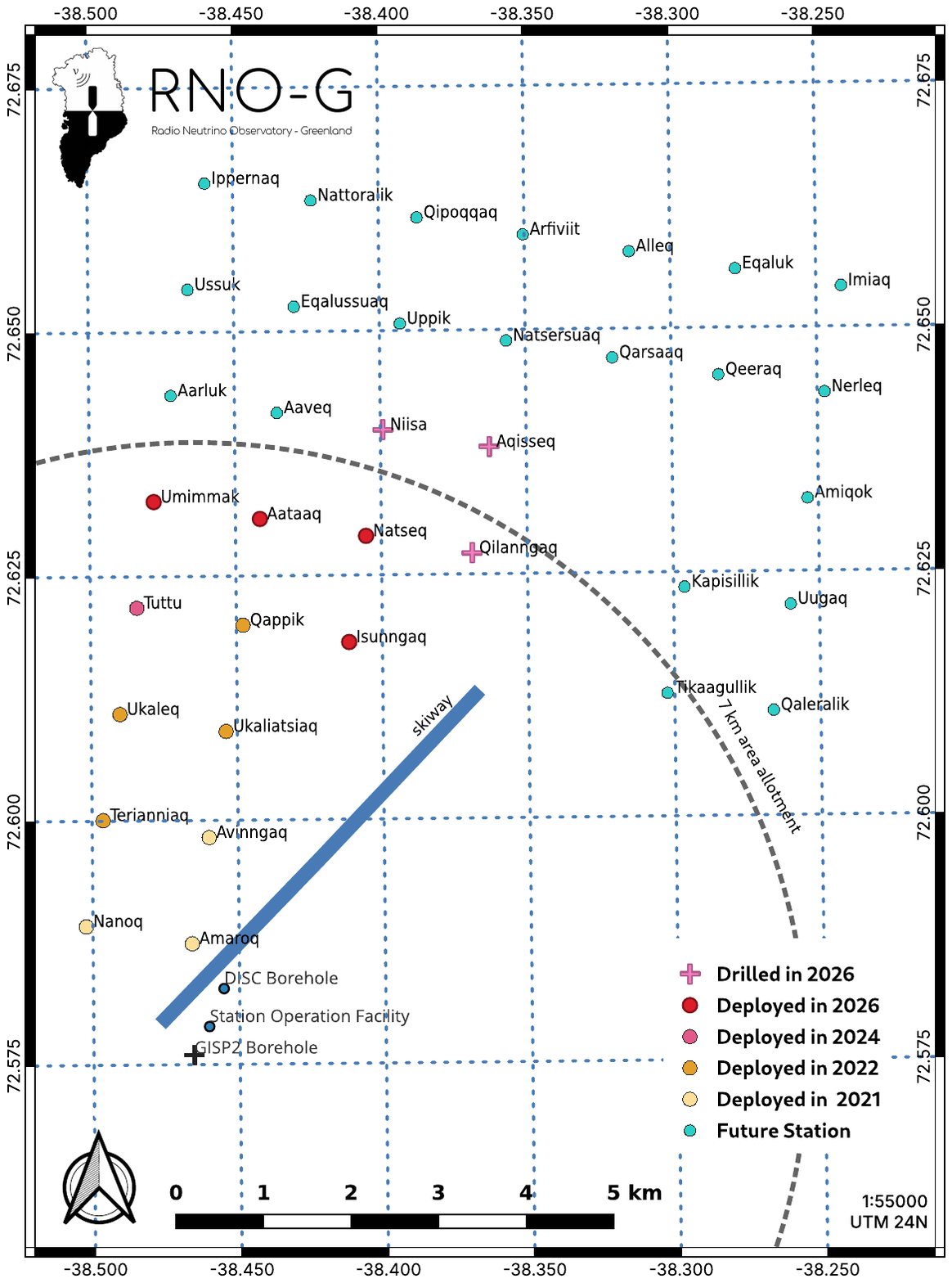}
         \caption{The RNO-G detector layout after the 2026 deployment season.}
         \label{fig:rnog}
\end{wrapfigure}

In the last decade, radio-based detectors have been established as one of the most promising approaches to measure the ultra-high-energy (UHE) cosmic neutrino flux \cite{Barwick:2022vqt}. Combined with ice as the detector medium, this technology can resource-efficiently monitor several cubic kilometers of effective detector volume. The sensitivity to UHE neutrinos derives from the coherent radio emission at the Cherenkov angle from particle showers in a dense medium called Askaryan emission \cite{Askaryan}. Due to the long attenuation lengths $\mathcal{O}(\qty{1}{\kilo\metre})$ in ice, the emitted radio waves are still detectable by sparsely instrumented radio antenna arrays installed near the surface of a glacier.

For such detectors, two effects directly impact the sensitivity: First, the density gradient in the top $\mathcal{O}(\qty{100}{\metre})$ of the glacier impacts the radio propagation by downward bending the trajectories of signals coming from below the surface. This can lead to signals missing the antennas too close to the surface for specific geometries. In this way, the deeper the antennas are installed (down to where this bending effect is minimal), the more effective ice volume can be monitored, increasing the sensitivity of the detector. Second, the more antennas can be installed in as little amount of time and with the minimal required resources, the easier it becomes to build a larger detector, directly increasing the sensitivity. Naturally, those two constraints are in direct competition with each other, as installing antennas deeper costs significantly more time and resources.

The Radio Neutrino Observatory in Greenland (RNO-G) \cite{RNOG}, currently under construction near Summit Station, is already the largest operating neutrino detector of its kind. When fully constructed, it will consist of 35 autonomously operating antenna clusters, called stations (See Fig. \ref{fig:rnog}), arranged on a square grid with \SI{1.25}{\kilo\meter} spacing. So far, 12 of these stations have been constructed. Each of these stations combines two approaches, navigating the previously described constraints. "Deep" antennas are installed in three, \SI{100}{\meter} deep boreholes and "shallow" LPDA antennas, currently buried by hand just below the surface. In this contribution, we will describe the signal chain of an RNO-G station and highlight a newly developed electrical melter to improve the deployment of the "shallow" antennas.

While still under construction, the electronics used by the sequentially built stations are continuously under revision. Here, we focus on the system design of the first 7 stations \cite{Agarwal_2025}. A single RNO-G station operates on a \SI{25}{\watt} power budget provided by two solar panels delivering up to \SI{350}{\watt}. In addition, several stations have since been retrofitted with micro wind turbines to provide power and extend the sensitivity of the detector during the Arctic winter.

The 15 antennas in the boreholes consist of eleven fat-dipole antennas (Vpols: \SIrange{140}{600}{\mega\hertz}) sensitive to vertically polarized signals and four quad-slotted cylinder antennas (Hpols: \SIrange{300}{850}{\mega\hertz}) sensitive to horizontally polarized signals. All of these antennas are installed with a front-end amplifier and fiber transceivers sending the signals to the surface. Their trigger consists of a four-Vpol phased array \cite{ALLISON2019112}. The nine surface antennas have so far been buried by digging trenches right below the surface to fit the larger Log Periodic Dipole Array (LPDA: \SIrange{100}{1000}{\mega\hertz}) antennas. They operate with a threshold-based coincidence trigger.

The DAQ system records signals at a sampling rate between 2.4 and 3.2 gigasamples per second and saves them to a micro-SD card at a rate of \SI{\sim 1}{\hertz} with a sample resolution of 12 bits. A low-power single-board computer is used to control both the RAdio DIgitizer and Auxiliary Neutrino Trigger (Radiant) board used for digitization and allowing for on-chip multi-event buffering and the FLexible Octal WavEform Recorder (Flower) board housing and FPGA used for the phased array trigger. All of these systems are installed at the surface, allowing for remote communication via LoRaWAN and LTE antennas to monitor the station health and update the firmware.

\section{Slot Melters for LPDA instrumentation}
\label{sec:slot melter}

To install the shallow LPDA antennas used by RNO-G in the Greenlandic ice sheet, the current method involves digging trenches deep enough to cover the full height of the antennas. With nine/eight LPDA antennas per station (the newer station designs were updated to only include eight instead of nine LPDA antennas), this means several hours of physical labor in freezing temperatures at an altitude higher than \SI{\sim 3}{\kilo\meter} above sea level. Apart from the significant physical effort, the trenches have to be wide enough for a person to fit in to reach the desired depth, resulting in a cavity significantly bigger than the width of the antennas. This means there is a large discrepancy between the location and orientation where the antenna is supposed to be installed and where it is actually installed. The discrepancy is in the order of \SI{\sim 20}{\centi\meter} positional and \SI{\sim 15}{\deg} rotational uncertainty. While more accurate position and orientation values can be obtained during post-calibration steps, non-symmetrical configurations in the antenna array can impact the sensitivity and reconstruction capabilities of the detector. Another issue with installing the antennas in dug trenches is the fragile tines of the LPDA antennas. When refilling the trenches with snow, especially the longer, upper tines can easily bend or be damaged, altering the antenna response. Furthermore, the accumulating snow pressure has been shown to push down the longer tines after installation. This method also leaves uncertainties about the snow properties (density, refractive index) surrounding the antenna affecting the antenna response that cannot easily be resolved.

\begin{figure}[ht]
         \centering
         \includegraphics[height=2.5in]{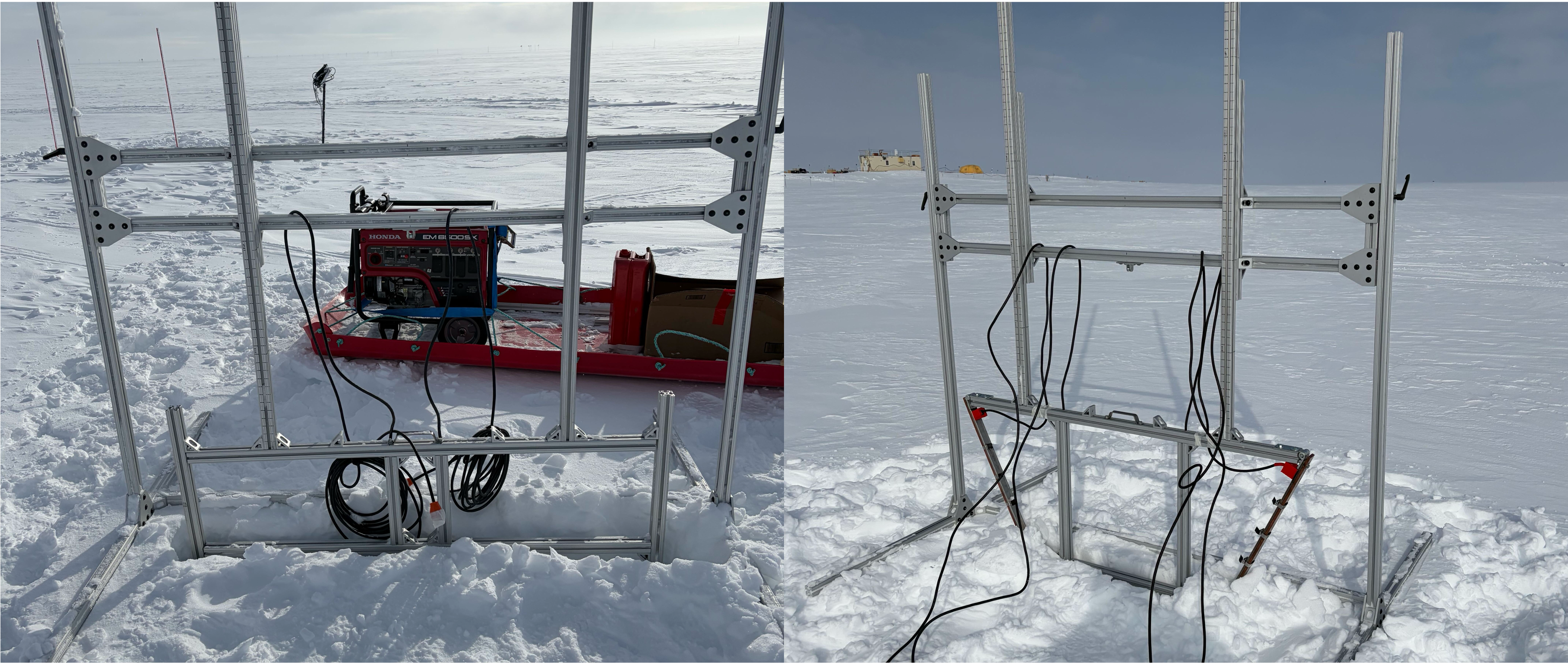}
         \caption{The two slot melter configurations with the V-shaped melter on the left and the H-shaped melter on the right.}
         \label{fig:melter}
\end{figure}

To address all of these issues, we designed, built, and tested an autonomous slot melter, able to create a precise cavity to deploy the LPDA into. The heat is provided by electrical strip heaters from the \textit{Tempco Electric Heater Corporation}. A single strip heater runs with \SI{1.5}{\kilo\watt} at \SI{120}{\volt} supplied by a \SI{6.5}{\kilo\watt} generator. To ensure uniform heating along the snow surface, the strip heaters were bolted onto a \SI{1.5}{in} (\SI{3.81}{\centi\meter}) copper bar. The verticality of the cavity is maintained by a heavy aluminum frame around the strip heaters guided by a rail system (see figure \ref{fig:melter}). As the electrical melter was designed to operate in an environment surrounded by snow and, therefore, water, the electrical connectors were engulfed in high-temperature epoxy after the cables were attached.

There are two ways of installing LPDA antennas for RNO-G: The downward-facing antennas for the main science case of detecting neutrinos, and the upward-facing antennas for the purpose of detecting cosmic rays. This means both installation types require different cavities to perfectly fit the antennas. To address this difference, two slot melter configurations were developed: A V-shaped or wedge-shaped melter attachment with two copper bars housing four strip heaters (\SI{6.0}{\kilo\watt}) in total to fit the downward-facing antennas, and a horizontal-shaped (H-shaped) melter with a single copper bar housing two strip heaters (\SI{3.0}{\kilo\watt}) in total to fit the upward-facing antennas.

\section{Field Tests in Greenland}
\label{sec:results}

To test how well the slot melter performed in the field, all components were shipped to Summit Station in Greenland for the field season during the summer of 2026. Two days were used for testing, and three slots were melted during this time. Here, several questions were answered, helping to evaluate its performance and feasibility for future use in a large-scale deployment campaign. 

\begin{wrapfigure}{r}{0.45\textwidth}
         \centering
         \includegraphics[height=3.9in]{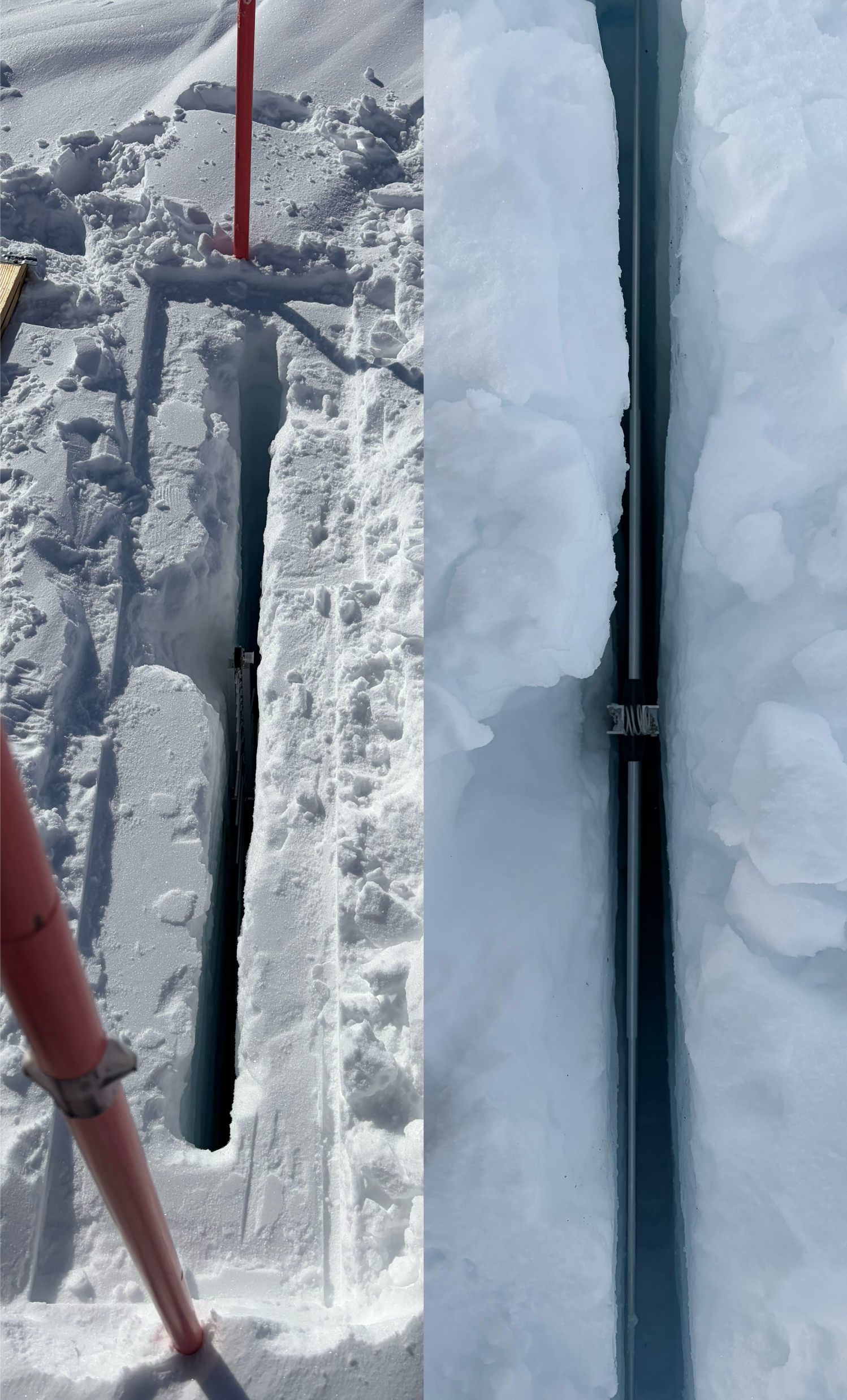}
         \caption{Upward (left) and downward (right) facing LPDA antenna installed in the melted slots.}
         \label{fig:slots}
\end{wrapfigure}

The first slot was a \SI{165}{\centi\meter} deep, H-shaped slot, melted in \SI{\sim 150}{\min} from the top of the snow surface. The melting process was completely autonomous, with regular depth and time measurements taken to later assess the melting speed (Fig. \ref{fig:slots}, left). It was noticed that the snow drift has the potential to freeze to the melting frame, stopping the autonomous melting process, but this can be easily prevented by blocking the slot from the direction of the wind. The melting speed was measured and approximated using a linear function to \SI{63.5}{\centi \meter \per \hour} after an initial time of \SI{\sim 20}{\min} to warm up to the operational temperature. 

The second slot was a \SI{165}{\centi\meter} deep, V-shaped slot, melted in \SI{\sim 110}{\min}. This slot was started from \SI{\sim 30}{\centi\meter} below the surface, which had been prepared by mechanical auger drills. The melting process was again fully autonomous, with sparse depth and time measurements (Fig. \ref{fig:slots}, right). Due to the staged mounting setup of the heating strips on the V-shaped melter, the melting was started with only the lower segments turned on, while the upper segments were turned on after \SI{\sim 20}{\min} of heating. This approach, together with the shape of the melter, ensures a non-linear melting speed until the full melter is engaged in the snow at \SI{\sim 150}{\centi\meter}. The melting speed was measured and approximated to \SI{70}{\centi \meter \per \hour}

\begin{figure}[ht]
         \centering
         \includegraphics[height=2.15in]{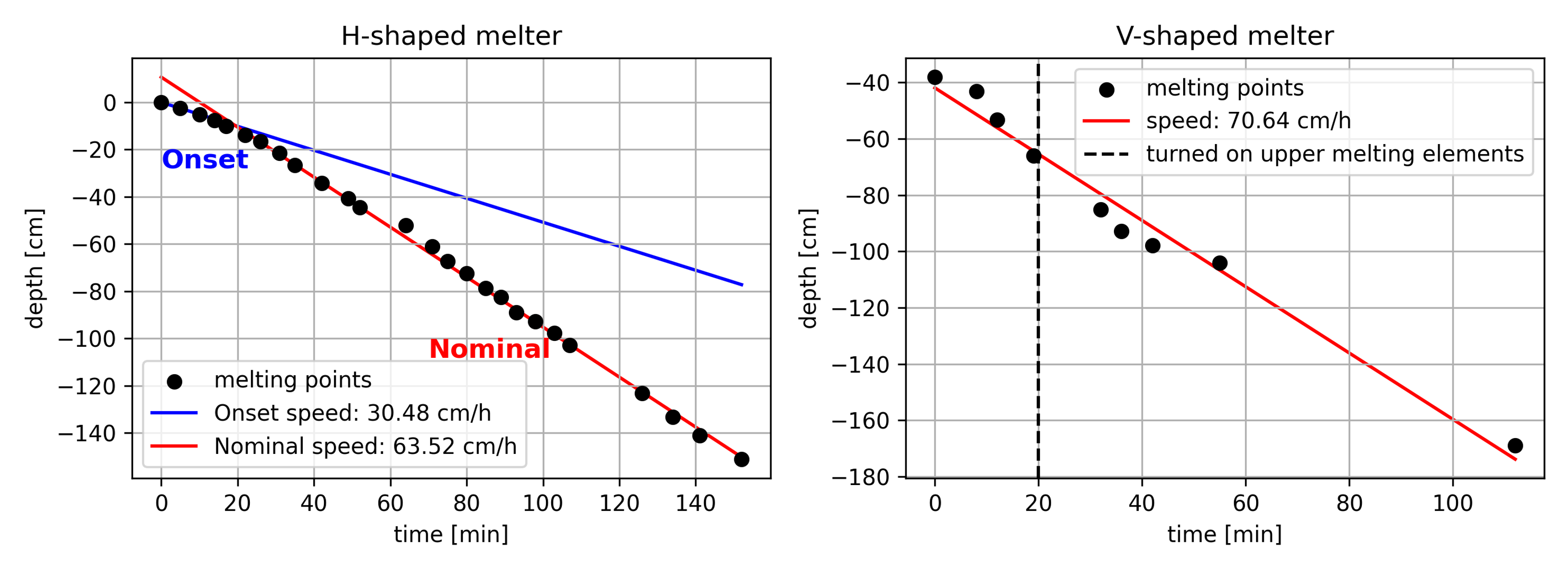}
         \caption{Melting speed of the H-shaped melter (left) and the V-shaped melter (right).}
         \label{fig:speed}
\end{figure}

The third slot was a \SI{445}{\centi\meter} deep, horizontal-shaped slot, melted in \SI{\sim 510}{\min} started from \SI{\sim 30}{\centi\meter} below the surface. Most of the melting process was autonomous. However, during the melting process, one of the extension beams, needed to reach deeper depths, created a gap in which the rail system got stuck. This created a window of about two hours during which the melter did not increase the depth, leading to an effective time of \SI{\sim 390}{\min} to reach the final depth. During nominal operations, the melting speed was \SI{59}{\centi \meter \per \hour} and, so far, appears to be independent of depth and the connected increased density.

To improve the melter in the future, longer extension beams are envisioned, allowing melting of \SI{\sim 200}{\centi\meter} deep slots without creating a gap that can potentially get stuck. For deeper slots, the connection between the extension beams is under revision. An additional issue noticed in the field was an overheating of the strip heaters leading to a melting of the epoxy, exposing the connectors, and risking a short circuit. This problem is currently under investigation, and we are developing improved, watertight strip heaters in cooperation with the supplier. The generators used to power the melter (Honda EM6500SX) ran continuously during the tests without issues, but had to be started in room temperature conditions. The two nominal slots to \SI{165}{\centi\meter} depth required six to seven liters of diesel for the melting process. 

The final cavities (Fig. \ref{fig:slots}) showed all the desired properties of a narrow slot, precisely shaped to the size of the antennas with \SI{\pm 2}{\centi\meter} in spatial uncertainty and \SI{\pm 1}{\deg} in rotational uncertainty. Furthermore, this technique makes it possible to cover the antenna-filled slot with a wooden cover and access it in future years to retrieve the antennas.

\section{Conclusion and Outlook}
\label{sec:conclusion}

For this contribution, we describe the system design of the first seven stations of RNO-G. In addition, we present a novel method to deploy LPDA antennas into Polar ice sheets with an electrical melter design. The new approach has been shown to significantly improve the accuracy of the antenna deployment while reducing the physical effort required. While some minor issues emerged during field tests in Greenland, the autonomous melting worked well and showed great potential to be used in future deployment campaigns. 
As RNO-G is still under construction with 23 stations left to build for the completed array, efficient antenna deployment remains a major logistical constraint. The presented melter can offer all the described advantages while requiring resources such as generators and fuel. The autonomous melting is estimated to make it possible for a "shallow-only" part of a station to be constructed and operationally ready within one day. Beyond RNO-G, planning for the radio array of IceCube Gen2 has begun, where several hundred of these stations are planned to be deployed near the South Pole. For such a large operation, efficient antenna deployment will also be a major consideration.


\bibliographystyle{JHEP}
\bibliography{references}


\section*{Full Author List: RNO-G Collaboration}
\scriptsize
\noindent
S.~Agarwal$^{1}$,
J.~A.~Aguilar$^{2}$,
N.~Alden$^{3}$,
S.~Ali$^{1}$,
P.~Allison$^{4}$,
M.~Betts$^{5,6}$,
D.~Besson$^{1}$,
A.~Bishop$^{7}$,
O.~Botner$^{8}$,
S.~Bouma$^{9}$,
S.~Buitink$^{10,11}$,
R.~Camphyn$^{2}$,
J.~Chan$^{7}$,
S.~Chiche$^{2}$,
B.~A.~Clark$^{12}$,
K.~Couberly$^{1}$,
D.~Dakroub$^{1}$,
K.~D.~de Vries$^{13}$,
C.~Deaconu$^{3}$,
P.~Giri$^{14}$,
C.~Glaser$^{8,15}$,
H.~Gui$^{4}$,
A.~Hallgren$^{8}$,
J.~C.~Hanson$^{16}$,
K.~Helbing$^{17}$,
B.~Hendricks$^{5,18}$,
J.~Henrichs$^{19,9}$,
N.~Heyer$^{8}$,
C.~Hornhuber$^{1}$,
E.~Huesca Santiago$^{19}$,
K.~Hughes$^{4}$,
A.~Jaitly$^{19,9}$,
A.~Karle$^{7}$,
J.~L.~Kelley$^{7}$,
C.~Kopper$^{9}$,
M.~Korntheuer$^{2,13}$,
M.~Kowalski$^{19,20}$,
I.~Kravchenko$^{14}$,
R.~Krebs$^{5,18}$,
M.~Kugelmeier$^{7}$,
D.~Kullgren$^{5,18}$,
R.~Lahmann$^{9}$,
C.-H. Liu$^{14}$,
Y.~Liu$^{4}$,
M.~J.~Marsee$^{21}$,
K.~Mulrey$^{11}$,
M.~Muzio$^{7}$,
A.~Nelles$^{19,9}$,
A.~Novikov$^{22}$,
A.~Nozdrina$^{4}$,
E.~Oberla$^{3}$,
N.~Punsuebsay$^{22}$,
M.~Ravn$^{8}$,
R.~Reimann$^{15}$,
Z.~Riesen$^{4}$,
A.~Rifaie$^{17}$,
D.~Ryckbosch$^{23}$,
Y.~Schaper$^{15}$,
F.~Schl{\"u}ter$^{2}$,
O.~Scholten$^{13,24}$,
P.~Schriefer$^{9}$,
D.~Seckel$^{22}$,
M.~F.~H.~Seikh$^{1}$,
Z.~S.~Selcuk$^{19,9}$,
J.~Stachurska$^{23}$,
J.~Stoffels$^{13}$,
S.~Toscano$^{2}$,
D.~Tosi$^{7}$,
J.~Tutt$^{6}$,
N.~van Eijndhoven$^{13}$,
A.~G.~Vieregg$^{3}$,
A.~Vijai$^{12}$,
H.~Warnhofer$^{19,9}$,
N.~Weitkemper$^{15}$,
C.~Welling$^{5,6}$,
D.~R.~Williams$^{21}$,
P.~Windischhofer$^{3,19}$,
S.~Wissel$^{5,6}$,
A.~Zink$^{9}$,

\noindent $^{1}$  University of Kansas, Dept.\ of Physics and Astronomy, Lawrence, KS 66045, USA  \\
$^{2}$  Universit\'e Libre de Bruxelles, Science Faculty CP230, B-1050 Brussels, Belgium  \\
$^{3}$  Dept.\ of Physics, Dept.\ of Astronomy \& Astrophysics, Enrico Fermi Inst., Kavli Inst.\ for Cosmological Physics, University of Chicago, Chicago, IL 60637, USA  \\
$^{4}$  Dept.\ of Physics, Center for Cosmology and AstroParticle Physics, Ohio State University, Columbus, OH 43210, USA  \\
$^{5}$  Dept.\ of Physics, Pennsylvania State University, University Park, PA 16802, USA  \\
$^{6}$  Dept.\ of Astronomy and Astrophysics, Pennsylvania State University, University Park, PA 16802, USA  \\
$^{7}$  Wisconsin IceCube Particle Astrophysics Center (WIPAC) and Dept.\ of Physics, University of Wisconsin-Madison, Madison, WI 53703,  USA  \\
$^{8}$  Uppsala University, Dept.\ of Physics and Astronomy, Uppsala, SE-752 37, Sweden  \\
$^{9}$  Erlangen Centre for Astroparticle Physics (ECAP), Friedrich-Alexander-University Erlangen-N\"urnberg, 91058 Erlangen, Germany  \\
$^{10}$  Vrije Universiteit Brussel, Astrophysical Institute, Pleinlaan 2, 1050 Brussels, Belgium  \\
$^{11}$  Dept.\ of Astrophysics/IMAPP, Radboud University, PO Box 9010, 6500 GL, The Netherlands  \\
$^{12}$  Dept.\ of Physics, University of Maryland, College Park, MD 20742, USA  \\
$^{13}$  Vrije Universiteit Brussel, Dienst ELEM, B-1050 Brussels, Belgium  \\
$^{14}$  Dept.\ of Physics and Astronomy, Univ.\ of Nebraska-Lincoln, NE, 68588, USA  \\
$^{15}$  Dept.\ of Physics, TU Dortmund University, Dortmund, Germany  \\
$^{16}$  Whittier College, Whittier, CA 90602, USA  \\
$^{17}$  Dept. of Physics, University of Wuppertal  D-42119 Wuppertal, Germany  \\
$^{18}$  Center for Multimessenger Astrophysics, Inst.\ of Gravitation and the Cosmos, Pennsylvania State University, University Park, PA 16802, USA  \\
$^{19}$  Deutsches Elektronen-Synchrotron DESY, Platanenallee 6, 15738 Zeuthen, Germany  \\
$^{20}$  Institut f\"ur Physik, Humboldt-Universit\"at zu Berlin, 12489 Berlin, Germany  \\
$^{21}$  Dept.\ of Physics and Astronomy, University of Alabama, Tuscaloosa, AL 35487, USA  \\
$^{22}$  Dept.\ of Physics and Astronomy, University of Delaware, Newark, DE 19716, USA  \\
$^{23}$  Ghent University, Dept.\ of Physics and Astronomy, B-9000 Gent, Belgium  \\
$^{24}$  Kapteyn Institute, University of Groningen, PO Box 800, 9700 AV, The Netherlands  \\

\subsection*{Acknowledgments}

\noindent
We are thankful to the support staff at Summit Station for making RNO-G possible. We also acknowledge our colleagues from the British Antarctic Survey for building and operating the BigRAID drill for our project.

\noindent We would like to acknowledge our home institutions and funding agencies for supporting the RNO-G work; in particular the Belgian Funds for Scientific Research (FRS-FNRS and FWO) and the FWO programme for International Research Infrastructure (IRI), the National Science Foundation (NSF Award IDs 2411590, 2514262, 2514201, 2514208, 2514206, and collaborative awards 2310122 through 2310129), and the IceCube EPSCoR Initiative (Award ID 2019597), the Helmholtz Association, the German Research Foundation (DFG, Grant 556092823), the Swedish Research Council (VR, Grants 2023-00156, 2021-05449), the University of Chicago Research Computing Center, and the European Union under the European Unions Horizon 2020 research and innovation programme (ERC, Pro-RNO-G No 101115122 and NuRadioOpt No 101116890).

\end{document}